\documentstyle[12pt,epsf,aps,preprint]{revtex}
\begin{document}

\def\etal{{\it et.~al}}
\def\vs{{\rm versus~}}

\draft

\setlength{\baselineskip}{.2in}

\title{The 2-Leg Hubbard Ladder:\\ Computational Studies of New Materials}

\author{\sl D.J.~Scalapino\thanks{djs@vulcan.physics.ucsb.edu}} 

\address{Department of Physics, University of California\\
Santa Barbara, California 93106}

\date{\today}

\maketitle

\begin{abstract}

\setlength{\baselineskip}{.2in}

Computational studies of basic models of strongly-correlated electron
systems can provide guidance in the search for new materials as well as
insight into the physical mechanisms responsible for their properties.
Here, we illustrate this by examining what numerical computations have shown
us about the 2-leg Hubbard ladder.
\end{abstract}

\newpage

This workshop in honor of Zachary Fisk's 60$^{\rm th}$ birthday is to explore 
the future
 of materials physics, an area to which he has contributed so much.  
It addresses the challenges and opportunities posed by complex materials.  
Here, 
in this session on computation, I want to focus on the problems of strongly 
interacting electronic systems, specifically the problem of the 2-leg
Hubbard ladder. While this is a special case, I believe it
illustrates the way in which numerical studies of simple models
can help in the search for new materials, such as the cuprate ladder
materials \cite{HT95,Ueh96}, as well as how such calculations can provide
 insight into more 
general problems 
such as that posed by the high
$T_c$-layered cuprates.  It also represents an example of a
case in which the interest in these materials arose from numerical
calculations, which suggested that an undoped ladder would have a spin gap
and a doped system would have power law pairing correlations \cite{DRS92}.

The Hamiltonian for the 2-leg Hubbard ladder is
\begin{equation}
H=-t\sum_{i,\lambda,s} \left(c^\dagger_{i+1, \lambda s} c_{i,\lambda s} +
c^\dagger_{i,\lambda s} c_{i+1, \lambda s}\right) - t_\perp \sum_{i, s}
\left(c^\dagger_{i,1s} c_{i, 2s}+ c^\dagger_{i,2s} c_{i, 1s}\right) + U \sum_{i, \lambda}
n_{i, \lambda\uparrow}n_{i, \lambda\downarrow}
\label{oneone}
\end{equation}
Here $c^\dagger_{i, \lambda s}$ creates an electron of spin $s$ at rung $i$
on leg $\lambda=1$ or 2.  The hopping along a leg is $t$ and across a rung
is $t_\perp$.  There is an onsite Coulomb interaction $U$ and the filling
is set by a chemical potential in the Monte Carlo simulations and by a
fixed electron number in the density matrix renormalization group (DMRG)
 calculations. 

The general properties of the 2-leg ladder \cite{DR96}, which have been 
found from both
numerical \cite{NSW96,NBSZ97} and analytic studies \cite{RGS93,BF96,Sch99},
 can be summarized as 
follows.  At
half-filling, the ground state has both a charge gap $\Delta_c$ and a spin
gap $\Delta_s$.  It is an insulator with short-range antiferromagnetic
correlations and spin gapped $S=1$ magnon excitations.  As $t_\perp/t$ 
increases the ground state of the half-filled
ladder adiabatically crosses over from a spin-gapped Mott
insulator to a band insulator \cite{NSW96} and $\Delta_s/\Delta_c$ goes 
to 1. For small values of $U$, this crossover to the band insulator
occurs when $t_\perp/t$ approaches 2.  This corresponds to the value at
which the bottom of the antibonding band rises above the top of the bonding
band and the non-interacting, half-filled  system becomes a Bloch insulator.  
When the
spin gapped Mott insulator is doped so that the site occupation $n=1-x$, 
the system goes into a Luther-Emery
phase \cite{LE74} with power law pair field and $4k_F$-CDW correlations. 
The sign of
the rung-rung pair field correlations are positive while the rung-leg
correlations are negative, implying ``$d$-wave-like'' pairing.
In the Luther-Emery phase, the power law decay of the pair field
$\ell^{-1/2\kappa}$ is conjugate to the $\ell^{-2\kappa}$ decay of the 
$4k_F$-CDW
correlations.  As one goes
towards the insulating state, Umklapp processes play an important role 
\cite{LHR00} and
Schulz \cite{Sch99} has shown that $\kappa\to1$ 
and the pairing
correlations are dominant.  Finally, we note that the superfluid density
vanishes as the doping $x$ goes to 0.

These properties make the 2-leg Hubbard ladder an interesting laboratory
for studying the effect of various parameters on the pairing.  In Figure 1,
a DMRG calculation \cite{NSW96} of
the rung-rung pair field correlation function
\begin{equation}
D(\ell)=\left\langle \Delta_{i+\ell} \Delta^\dagger_i\right\rangle
\label{onetwo}
\end{equation}
is shown for a $2\times 32$ ladder with a hole doping of $x=0.125$,
$U/t=8$, and different values of $t_\perp/t$. Here
\begin{equation}
\Delta^\dagger_i= \left(c^\dagger_{i, 1\uparrow} c^\dagger_{i, 2\downarrow}
- c^\dagger_{i, 1\downarrow} c^\dagger_{i, 2\uparrow}\right)
\label{onethree}
\end{equation}
creates a singlet pair on the $i^{\rm th}$ rung. As $t_\perp/t$ increases,
the pairing correlations initially increase.  However, as shown for 
$t_\perp/t=2$, when $t_\perp/t$ becomes too large,  we
find that the pairing correlations are suppressed.  A closer look at this
behavior is shown in Fig.~2a.  Here an average strength of the rung-rung
pairing correlations
\begin{equation}
\bar D= \frac{1}{5} \sum^{12}_{\ell=8} D(\ell)
\label{onefour}
\end{equation}
is shown \vs $t_\perp/t$ for $U/t=8$ at several dopings. In Fig.~2b,
$\bar D$ \vs $t_\perp/t$ is shown for various values of $U/t$ at a doping
$x=.0625$. Separate calculations of the pair gap show that it has the
same behavior as $\bar D$ as a function of $t_\perp/t$ and $x$.  
The strength of the pair binding
increases as $x$ decreases and the Mott insulating state is approached
while, as noted previously, the superfluid density decreases as $x$ goes
to zero.  The pairing also depends upon $U$, and as shown in Fig.~2b,
peaks for $U$ of order the bandwidth. For values $U$ large 
compared with
the bandwidth, the effective exchange interaction varies as $U^{-1}$ and
both the insulating spin gap and the pairing strength decrease as $U$
increases.   

The peak in $\bar D$ as a function
of $t_\perp/t$ can be understood in terms of the behavior of the
single-particle spectral weight $A(p,\omega)$.  Results for $A(p,\omega)$
obtained from a maximum entropy analytic continuation of Monte Carlo data
\cite{NBSZ97,DS97} for a 2-leg ladder at $T=.25t$ with $U/t=4$, $t_\perp/t=1.5$,
and $\langle n\rangle=0.875$ are shown in Fig.~3.  Here, the solid curves
are for the bonding band ($p_y=0$) and the dotted curves are for the
antibonding band ($p_y=\pi$).  One sees that for this value of $t_\perp/t$,
the bonding band has spectral weight near the fermi level for $p\simeq
(\pi, 0)$ while the antibonding band has its spectral weight near the fermi
level for $p\simeq (0, \pi)$.  These Fermi points can be connected by
scatterings involving a large momentum transfer where, as we will see,
 the effective pairing
interaction $V$ is strong. Furthermore, the peak in both the
bonding and antibonding spectral weight is seen to disperse very slowly
near the bonding and antibonding fermi points leading to a large density of
states at the fermi energy.  For a value of $t_\perp/t \sim 1.6$ which is
slightly larger, the dispersion becomes even flatter and $\bar D$ peaks.  At
still larger values of $t_\perp/t$, the antibonding band pulls away from the 
Fermi energy leaving only the bonding
band with spectral weight near the Fermi level. In this case, the doped system 
behaves like a one-band Luttinger liquid and the pairing correlations vanish.
 
These computations suggest that an array of weakly coupled ladders should
exhibit superconductivity and that $T_c$ can be enhanced by adjusting the
ratio of the effective rung-to-leg hopping as well as optimally doping the
system. Additional calculations \cite{DSW00}
 have shown that the pairing can be further
enhanced if an exchange interaction $J_\perp \vec S_{i1} \cdot
\vec S_{i2}$ is added.  In a $CuO_2$ model, such an additional exchange
term can be seen to arise from intermediate states in which the two
exchanged electrons virtually occupy the oxygen site rather than one of the 
$Cu$ sites.  Chemically, one might try to enhance the rung hopping
and the rung exchange by substitutions off the ladder plane that change
the Madelung energy of the rung O. A more radical approach would be to
replace the rung O by another element such as $S$.
Clearly, the 2-leg ladder materials
represent an important area of new materials that merit further study.

Beyond this, however, computational studies of the 2-leg
ladder have provided insight into the more general problem of the high
$T_c$ cuprates.  Using Monte Carlo simulations \cite{DS97}, one can extract 
the effective pairing interaction $V (q, \omega_m)$. While these
calculations for a doped ladder were restricted to temperature $T\geq
0.25t$ because of the ``fermion sign'' problem, they showed that the momentum,
Matsubara frequency and temperature dependence of $V (q,
\omega_m)$ was remarkably similar to that of the spin susceptibility 
\begin{equation}
\chi(q,\omega_m) = \frac{T}{N}\, \int^\beta_0\, d\tau e^{-i\omega_m\tau}
\sum_q e^{i\vec q\cdot\vec \ell} \left\langle M^z_{i+\ell}(\tau) M^z_i
(0)\right\rangle
\label{onefive}
\end{equation}
Here $M^z_i=\left(n_{i\uparrow}-n_{i\downarrow}\right)/2$.  A comparison of
$V (q, \omega_m=0)$ \vs $q$ with
$\chi(q,\omega_m=0)$ for various temperatures is shown in Figures 4a and b.
As the temperature decreases and the short-range antiferromagnetic
correlations evolve in $\chi(q)$, the effective interaction grows in
strength and a broad peak appears at large momentum transfers.
It is this interaction which
leads to the strong scattering of pairs from regions near the bonding fermi
points $(p_{Fb}, 0)$ and $(-p_{Fb}, 0)$ 
to regions near the antibonding fermi points $(p_{Fa}, \pi)$ and $(-p_{Fa}, 
-\pi)$ and hence to the pair binding.

It is also useful to consider a strong-coupling local picture of the
pairing.  Figure 5 shows the results of a DMRG calculation of the amplitude
for removing 2 electrons \cite{SW00}
\begin{equation}
\left\langle N-2\left|\left(c_{i\uparrow}c_{j\downarrow}-
c_{i\downarrow} c_{j\uparrow}\right)\right| N\right\rangle
\label{onesix}
\end{equation}
from a half-filled 2-leg ladder.  The phase of this amplitude has been 
chosen so that the rung amplitude is positive.  Here one sees a clear
``$d_{x^2-y^2}$-like'' structure and from the size of the amplitude, one can
judge that the pair is extended over about 10 sites.  This amplitude is 
probing a
property of the half-filled ground state and although the superfluid
density for the undoped ladder vanishes, this spin-gapped Mott state
contains latent $d_{x^2-y^2}$-like pairs. To illustrate this,
consider a $2\times 2$ segment of a ladder in which the 4-sites are 
numbered in a clockwise manner. In a ladder 
with short-range antiferromagnetic correlations, the half-filled, 4-site
 cluster has a ground state \cite{ST97}
\begin{equation}
\left|\psi\rangle= N_0 \left(\Delta^\dagger_{14} \Delta^\dagger_{23} -
\Delta^\dagger_{12} \Delta^\dagger_{34}\right)\right|0\rangle
\label{oneseven}
\end{equation}
with
$\Delta^\dagger_{ij}=c^\dagger_{i\uparrow}c^\dagger_{j\downarrow} -
c^\dagger_{i\downarrow}c^\dagger_{j\uparrow}$.  This superposition of
valence bonds \cite{Ref} has $d_{x^2-y^2}$-like symmetry. 
The power law
quasi-long-range-order superconducting pairing phase arises when these
latent pairs are ``freed'' by the doping.

In summary, the dependence of the pairing correlations on the ratio
of the rung-to-leg hopping parameters $t_\perp/t$, the strength of the
Coulomb
interaction $U/t$ and the doping $x$ provide insight into the search for new
ladder materials that could exhibit enhanced superconductivity.
The effective pairing interaction $V (q,\omega_m)$ 
shown
in Fig.~4a and the resonating valence bond structure shown in Fig.~5,
illustrate the momentum space and real space nature of the pairing process
for the 2-leg Hubbard ladder.  The short coherence length and the relatively
high energy scale set by the gap magnitude of the 2D layered high $T_c$ cuprates suggest that the basis pairing mechanism in these materials involves 
short-range physics.  Thus we expect that what has been learned about the 
pairing process in the 2-leg ladder will also be relevant to the layered
cuprates.  

\acknowledgments

Pieces of this work were carried out with N~Bulut, T.~Dahm, R.~Noack, and 
S.R.~White.  This research was supported in part by the National Science
Foundation under Grant No.~DMR98-17242 and by the Department of Energy
under Grant No.~DE-FG03-85ER45197. A number of numerical computations were
carried out at the San Diego Supercomputer Center and we are grateful to them
for their support.

\begin{figure}
\begin{center}
{\epsfysize=3in \epsfbox{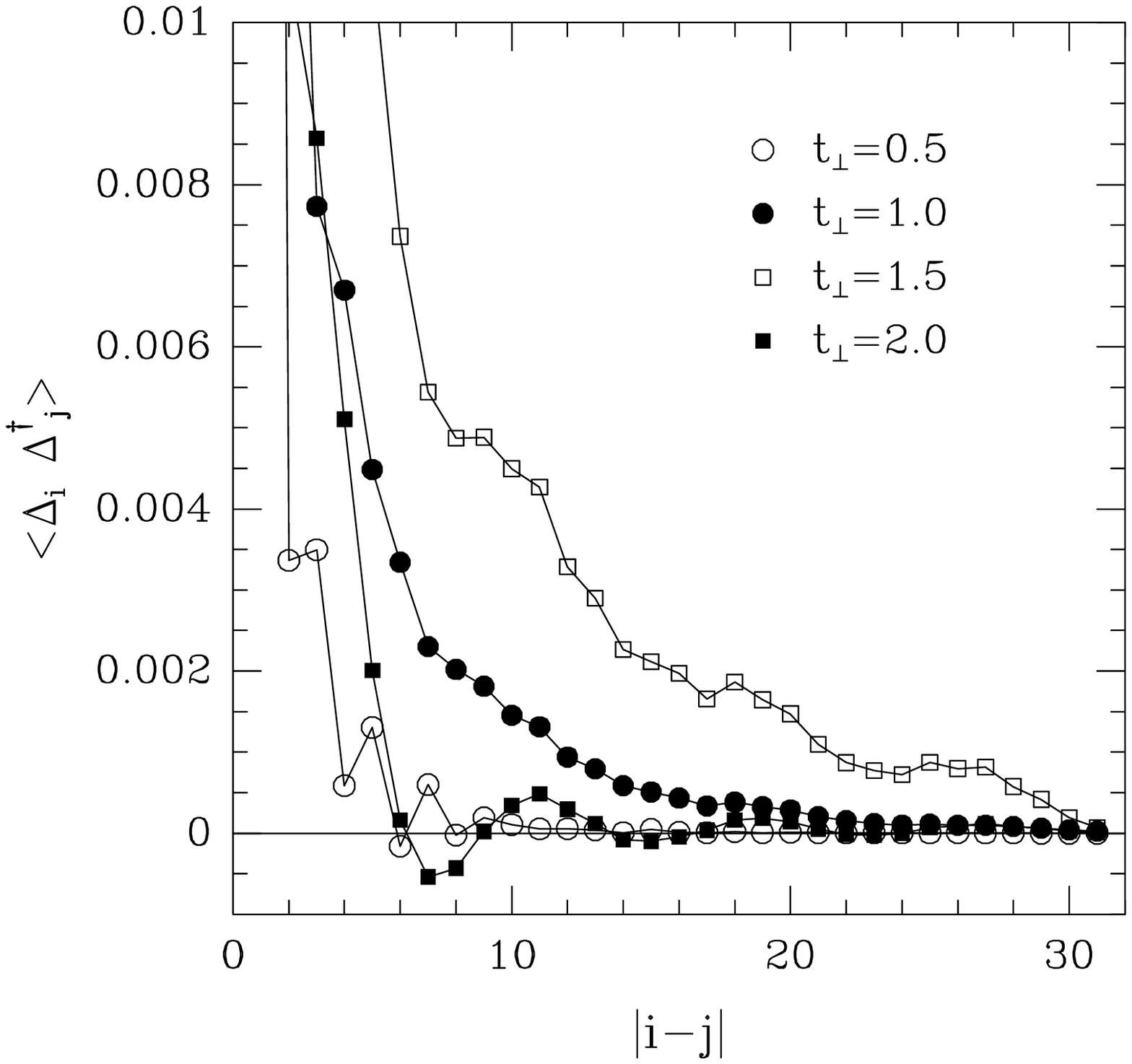}}
\end{center}
\caption{
The equal-time pair-field correlation function
$\langle\Delta_i \Delta^\dagger_j\rangle$ \vs $|i-j|$ for
$\langle n\rangle=0.875$, $U/t=8$ and various values of $t_\perp/t$.
}
\end{figure}

\begin{figure}
\begin{center}
{\epsfysize=3in \epsfbox{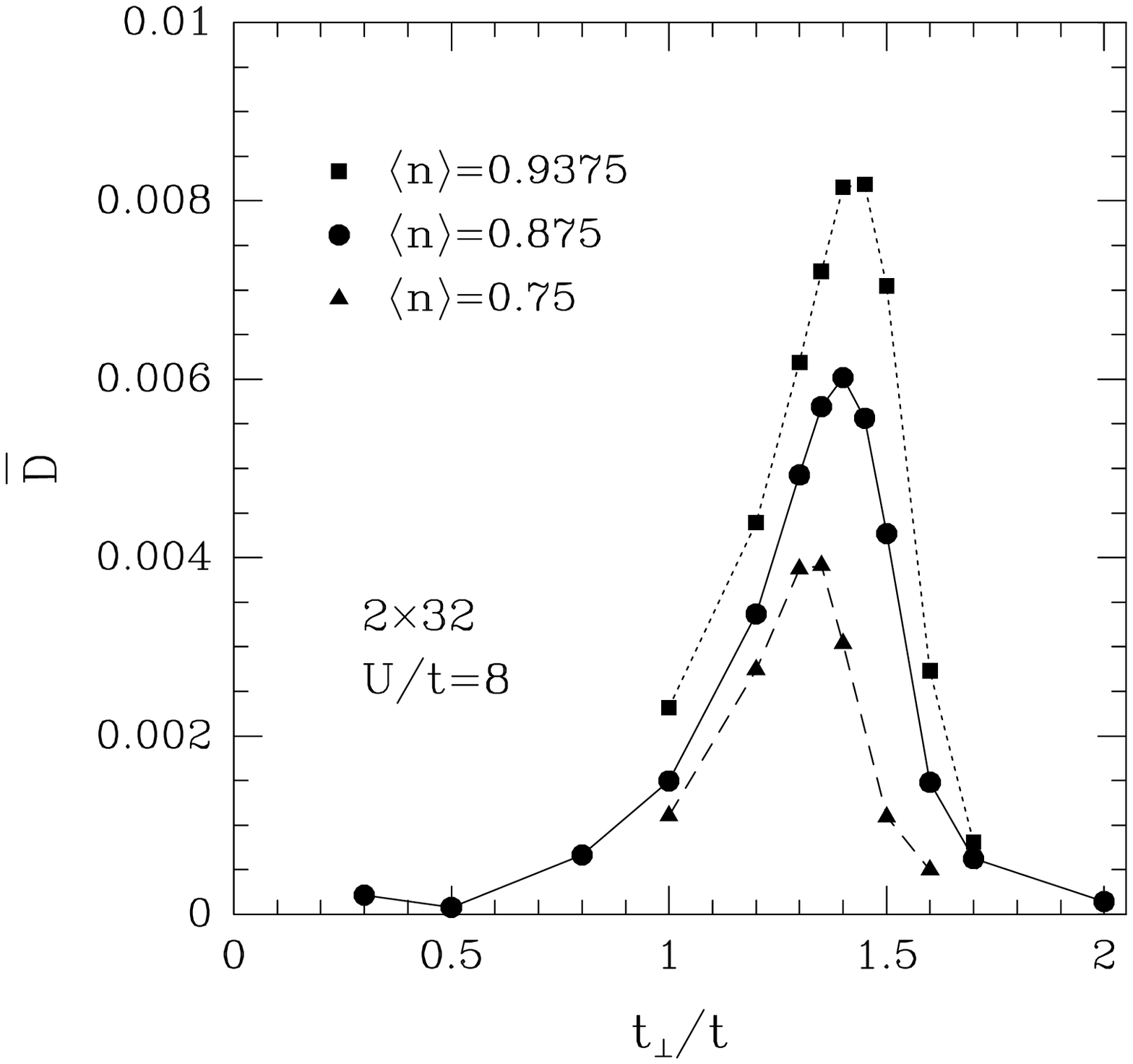}}
{\epsfysize=3in \epsfbox{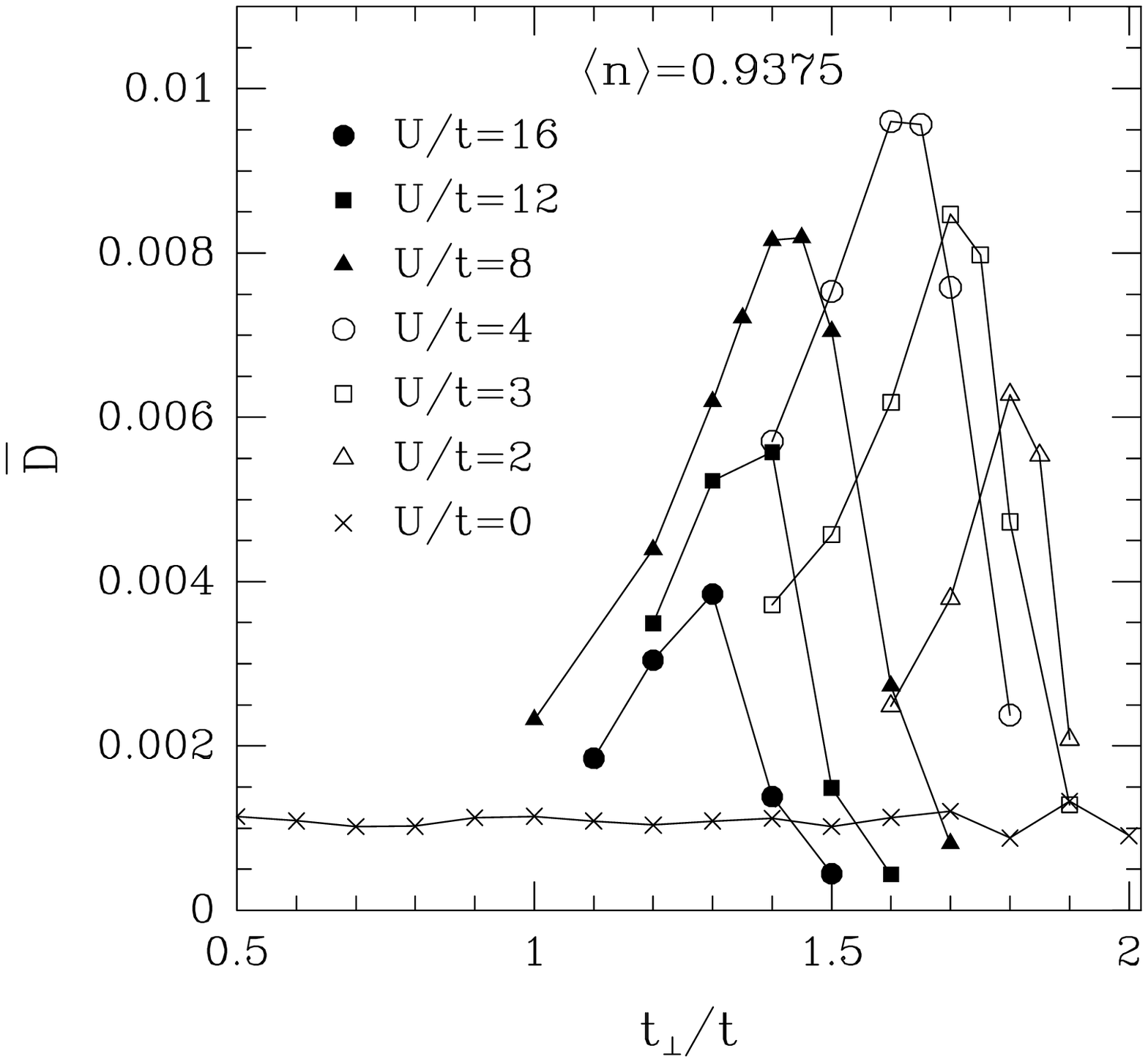}}
\end{center}
\caption{a) $\bar D$ versus $t_\perp/t$ for $U/t=8$ at
fillings $\langle n
\rangle=0.75$, 0.875, and 0.9375.
b)  $\bar D$ versus $t_\perp/t$ for various values of
$U/t$ at filling
$\langle n\rangle=0.9375$.
}
\end{figure}

\begin{figure}
\begin{center}
{\epsfysize=3in \epsfbox{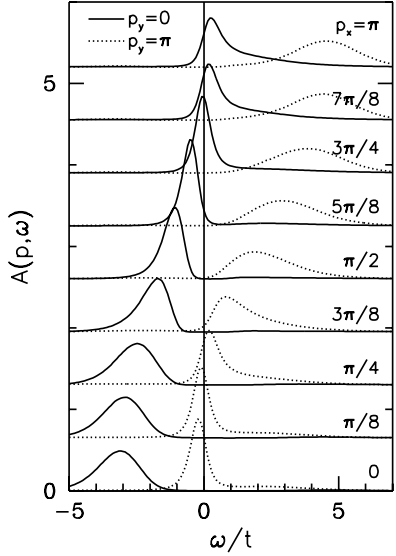}}
\end{center}
\caption{Single-particle spectral weight $A({\bf p}, \omega)$
versus $\omega$
for $t_\perp/t=1.5$, $T=0.25t$, $U/t=4$, and $\langle n\rangle=0.875$.  The
solid curves denote the results for the bonding band $(p_y=0)$ and the
dotted curves denote the results for the antibonding band $(p_y=\pi)$.
}
\end{figure}

\begin{figure}
\begin{center}
{\epsfysize=3in \epsfbox{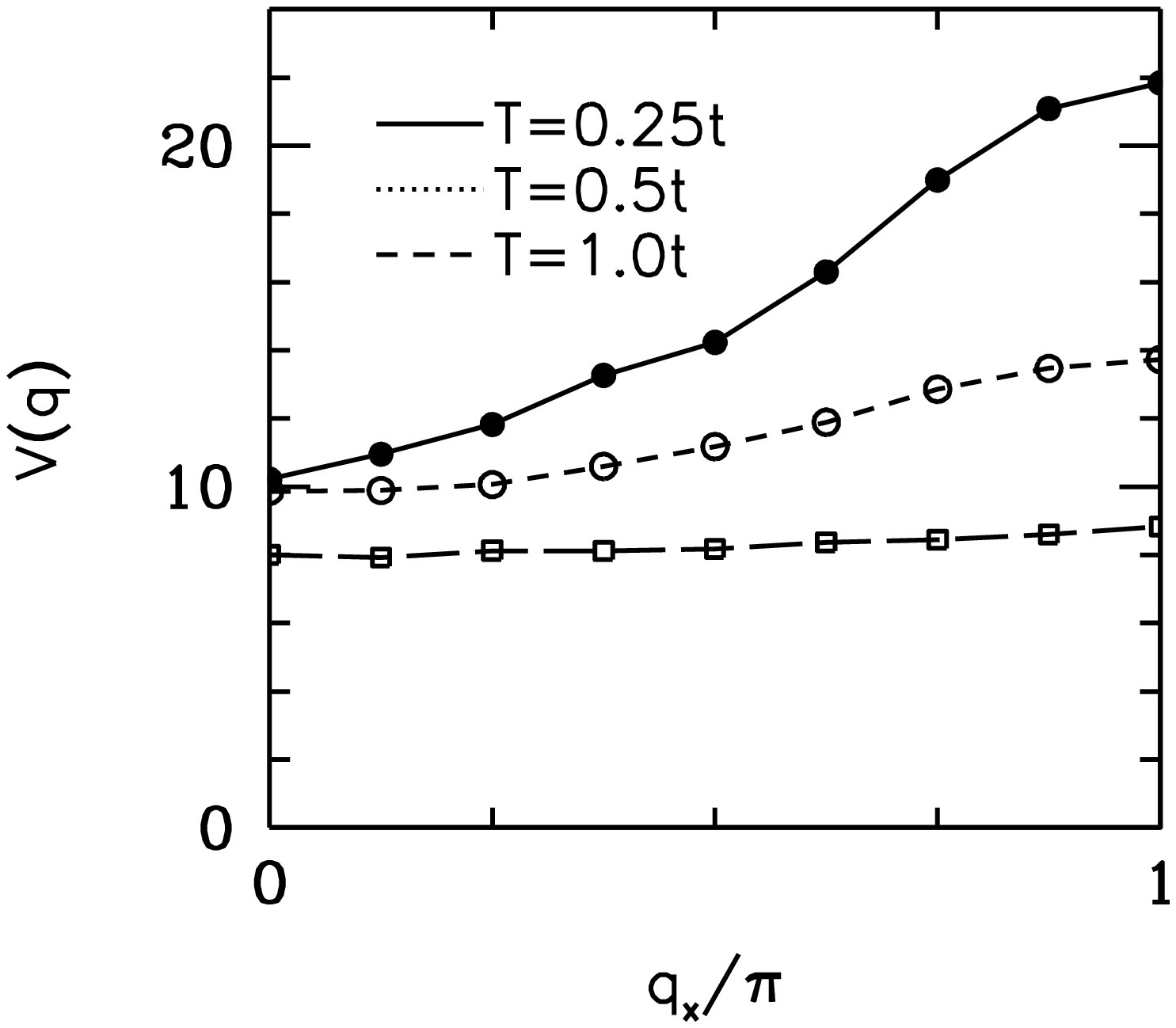}}
{\epsfysize=3in \epsfbox{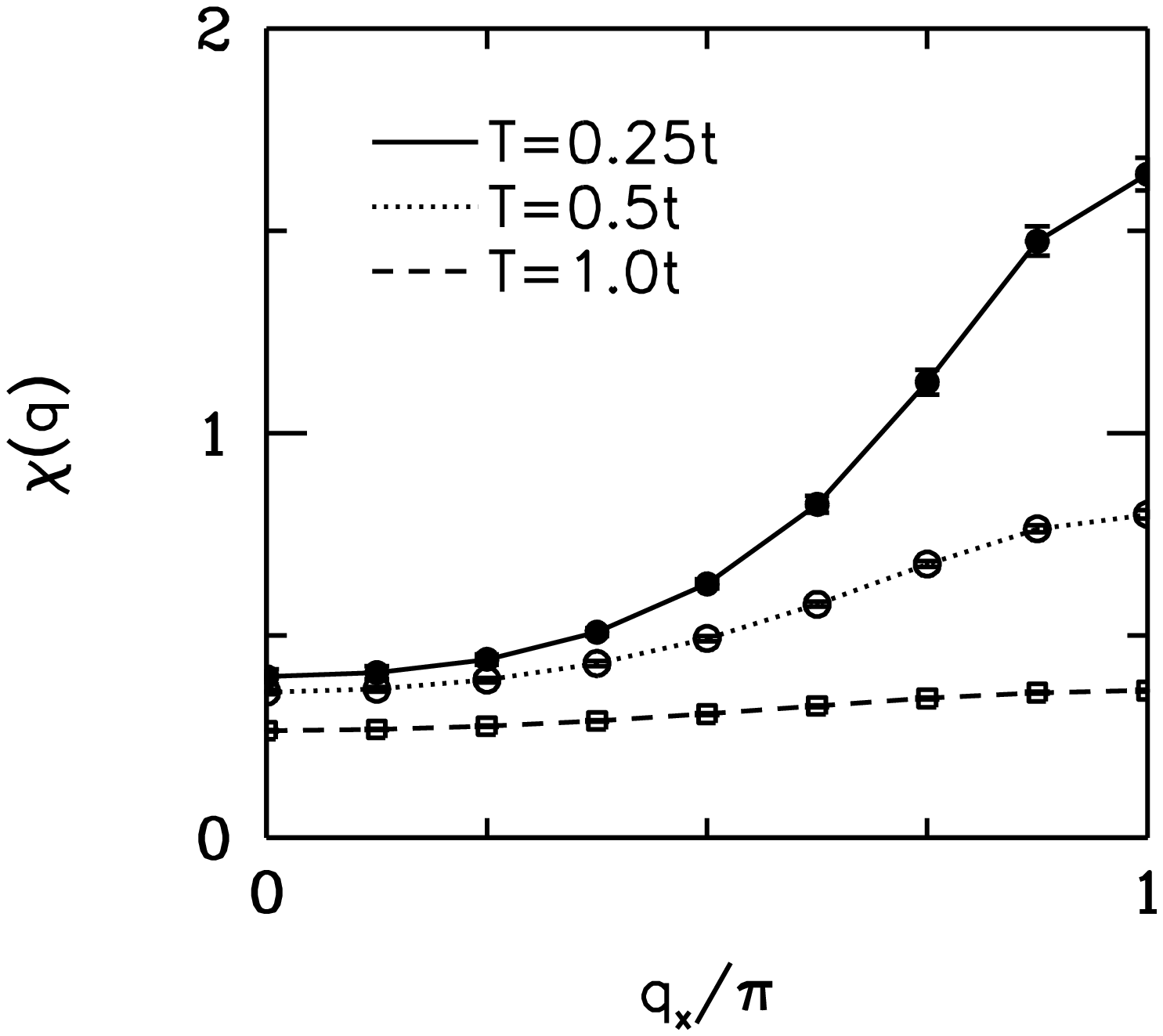}}
\end{center}
\caption{a) Momentum-dependence of the effective interaction
$V({\bf q})$
for $U=4t$, $\langle n \rangle=0.875$, and $t_\perp=1.5t$.  Here $V({\bf q})$
is measured in units of $t$, $q_y=\pi$ and $V({\bf q})$ is plotted as a
function of $q_x$. b) Momentum-dependence of the magnetic
susceptibility $\chi({\bf q})$
for $U=4t$, $\langle n \rangle=0.875$, and $t_\perp=1.5t$.  Here, $q_y=\pi$ and
 $\chi({\bf q})$ is plotted as a function of $q_x$.
}
\end{figure}

\newpage

\begin{figure}
\begin{center}
{\epsfysize=1.5in \epsfbox[146 318 463 477]{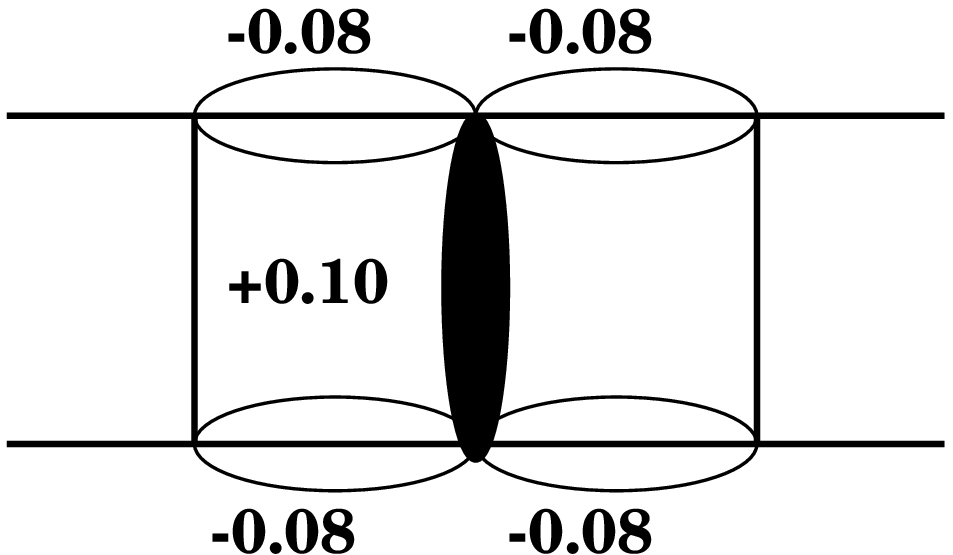}}
\end{center}
\caption{Schematic drawing of the pair-wave function showing
the values of
the off-diagonal matrix element  $\Bigl\langle
N-2\left|\left(c_{i\uparrow}c_{j\downarrow} -
c_{i\downarrow} c_{j\uparrow}\right)\right|N \Bigr\rangle$ for
removing a singlet pair between near-neighbor sites.
}
\end{figure}


\begin{references}

\setlength{\baselineskip}{.2in}

\bibitem{HT95} Z.~Hiroi and M.~Takano, {\sl Nature} {\bf 377} (1995) 41.

\bibitem{Ueh96} M.~Uehara \etal, {\sl J.~Phys.~Soc.~Jpn} {\bf 65} (1996)
2764.

\bibitem{DRS92} E.~Dagotto, J.~Riera, and D.J.~Scalapino, {\sl
Phys.~Rev.~B} {\bf 45} (1992) 5744.

\bibitem{DR96} E.~Dagotto and T.M.~Rice, {\sl Science} {\bf 271} (1996)
618.

\bibitem{NSW96} R.M.~Noack, D.J.~Scalapino, and S.R.~White, {\sl
Phil.~Mag.~B} {\bf 74} (1996) 485.

\bibitem{NBSZ97} R.M.~Noack, N.~Bulut, D.J.~Scalapino, and M.G.~Zacher,
{\sl Phys.~Rev.~B} {\bf 56} (1997) 7162.

\bibitem{RGS93} T.M.~Rice, S.~Gopalan, and M.~Sigrist, {\sl
Europhys.~Lett.} {\bf 23} (1993) 445; S.~Gopalan, T.M.~Rice, and
M.~Sigrist, {\sl Phys.~Rev.~B} {\bf 49} (1994) 8901.

\bibitem{BF96} L.~Balents and M.P.A.~Fisher, {\sl Phys.~Rev.~B} {\bf 53}
(1996) 12133.

\bibitem{Sch99} H.J.S.~Schulz, {\sl Phys.~Rev.~B} {\bf 59} (1999) R2471.

\bibitem{LE74} A.~Luther and V.J.~Emery, {\sl Phys.~Rev.~Lett.} {\bf 33}
(1974) 589.

\bibitem{LHR00} U.~Ledermann, K.~LeHur, and T.M.~Rice, {\sl Phys.~Rev.~B} 
{\bf 62} (2000) 16383.
 
\bibitem{DS97} T.~Dahm and D.J.~Scalapino, {\sl Physica C} {\bf  288}
(1997) 33.

\bibitem{DSW00} S.~Daul, D.J.~Scalapino, and S.R.~White, {\sl Phys.~Rev.~Lett.} {\bf 84} (2000) 4188.

\bibitem{SW00} D.J.~Scalapino and S.R.~White, {\sl Physica C} {\bf 341-348}
(2000) 367.

\bibitem{ST97} D.J.~Scalapino and S.A.~Trugman, {\sl Phil.~Mag.~B} {\bf 74} 
(1996) 607.

\bibitem{Ref} It would be natural to call this a short-range RVB state 
[P.W.~Anderson, {\sl Science} {\bf 235} (1987) 1169; S.A.~Kivelson, 
D.S.~Rokhsar, and J.P.~Sethna, {\sl Phys.~Rev.~B} {\bf 35} (1987) 8865.] 
except that the term RVB carries a variety of special connotations with it 
such as spinons. As noted, a half-filled ladder has only $S=1$ spin-gapped 
magnon excitations and is adiabatically connected to a band insulator.

\end{references}
\end{document}